\documentclass{mem}
\usepackage{natbib}\usepackage{txfonts}\usepackage{balance}
\usepackage{graphicx}
\usepackage[a4paper,breaklinks,dvipdfm]{hyperref}
\idline{75}{282}

\newcommand{\nh}{N$_{\rm H}$}

\newcommand{\ergscm}{\,erg\,cm$^{-2}$\,s$^{-1}$}
\newcommand{\Fx}{$\rm F_{\rm X}$}
\newcommand{\ergs}{\,erg\,s$^{-1}$}
\newcommand{\kms}{km s$^{-1}$}

\newcommand{\Lx}{$\rm L_{\rm X}$}
\newcommand{\fxfo}{F$_{\rm X}$/F$_{\rm opt}$}

\begin{document}
\def\teff{$T\rm_{eff }$}
\def\kms{$\mathrm {km s}^{-1}$}

\title{
Multi-frequency Studies of \\ Galactic X-ray Sources Populations
}

   \subtitle{Hard X-ray Galactic sources of low to intermediate \Lx \\ A search for isolated accreting black holes}

\author{
C. \,Motch\inst{1} 
\and
M. W. Pakull\inst{1}
          }

  \offprints{C. Motch}

\institute{
Observatoire Astronomique de Strasbourg, UMR 7550 Universit\'e de Strasbourg - CNRS, 11, rue de l'Universit\'e, F-67000 Strasbourg, France
\email{christian.motch@unistra.fr}
}

\authorrunning{C. Motch \& M. Pakull}

\titlerunning{Galactic X-ray Sources Populations}

\abstract{Our Galaxy harbours a large population of X-ray sources of intermediate to low X-ray luminosity (typically \Lx\ from 10$^{27}$ to 10$^{34}$\ergs). At energies below 2\,keV, active coronae completely dominate the X-ray landscape. However, the nature and the properties of Galactic sources detected at energies $\ga$ 2\,keV is much less constrained. Optical follow-up spectroscopic observations show that in addition to cataclysmic variables (CVs) and very active stellar coronae, massive stars (colliding wind binaries, quiescent high-mass X-ray binaries and $\gamma$-Cas analogs) account for a sizable fraction of the Galactic hard X-ray sources at medium flux (\Fx\ $\ga$ 10$^{-13}$\ergscm). Cross-correlations of the 2XMM-DR3 catalogue with 2MASS and GLIMPSE confirm the presence above 2\,keV of a large population of coronally active binaries, probably of the BY Dra and RS CVn types, in addition to many distant and absorbed massive stars. We also report the results of a specific optical identification campaign aimed at studying the nature of the optically faint hard X-ray sources and at constraining the surface density of black holes (BHs), either isolated and accreting from the interstellar medium or in quiescent binaries. Not astonishingly, most of our sample of 14 optically faint and X-ray hard sources are identified with CVs and Me stars. We do not find any likely counterpart in only three cases. Our observations also allow us to put an upper limit of 0.2 BH deg$^{-2}$ at \Fx = 1.3 10$^{-13}$\ergscm\ in directions toward the centre of our Galaxy. This implies a combined Bondi-Hoyle and $\dot{\rm M}$  to \Lx\ efficiency of accretion onto black holes of less than 10$^{-3}$.

}
\maketitle{}

\section{Introduction}

Our Galaxy presents a rich and contrasted X-ray landscape. Many different classes of Galactic objects can be the source of high energy emission and cover a wide range of spectral properties and X-ray luminosities. Active stellar coronae are by far the most numerous soft (kT\,$\sim$\,0.5\,keV) X-ray sources encountered at low Galactic latitudes. Their X-ray luminosities are in the range of 10$^{27}$ to 10$^{31}$\ergs. With \Lx\ typically higher than 10$^{35}$\ergs, classical High and Low-Mass X-ray binaries (HMXBs, LMXBs) occupy the bright end of the Galactic X-ray luminosity function. However, X-ray surveys have also shown the presence of a large population of relatively hard Galactic sources with \Lx\ in the range of 10$^{31}$ to a few 10$^{34}$\ergs. The nature and overall properties of these intermediate X-ray luminosity sources remains badly understood. Because of their faintness, these populations can only be studied in our Galaxy. Although CVs and coronally active binaries are likely to account for a large part of this population, many other kinds of objects have also been identified in this range of \Lx, namely X-ray binary transients in quiescent states, magnetic OB stars or colliding wind massive binaries. Among the most interesting objects that could appear in this \Lx\ regime are isolated compact remnants of early stellar formation accreting from the interstellar medium and low luminosity stages predicted by binary evolution models (e.g., Be + white dwarf, wind accreting neutron star binaries or precursors of LMXBs). 

Low-luminosity hard X-ray sources are also likely to be important contributors to the Galactic ridge X-ray emission (GXRE; see e.g. \citealt{worrall1982}). The nature of this extended and apparently diffuse emission remains debated. Its X-ray spectrum displays a prominent emission line at 6.7\,keV and resembles that of a thin thermal plasma with temperatures of 5-10\,keV \citep{koyama1986}. However, such a hot interstellar medium cannot remain bound in the Galactic gravitational potential well and its presence would require that unrealistically powerful sources of hot plasma concur to replenish it continuously. The close similarity of the spatial distribution of the GXRE with near-infrared emission \citep{rev2006} also favours an explanation in terms of unresolved emission of many low-\Lx\ sources. A deep Chandra observation of one of the brightest ridge emission close to Galactic Centre resolved at least 80\% of the diffuse emission into point sources at energies 6-7\,keV \citep{rev2009}. However, \cite{ebisawa2005} reach opposite conclusions based on another deep Chandra observation at $l$\,$\sim$\,28.5\degr.

The goal of this paper is to explore the nature of low to intermediate X-ray luminosity sources encountered in the Galaxy and shining in the hard (2-12\,keV) range.  Our work is based on spectroscopic identifications obtained at the telescope and identifications derived from the cross-correlation of XMM-Newton serendipitous sources with large optical and infra-red archival catalogues.

\section{The X-ray content of the XMM-Newton Galactic Plane Survey}

The Survey Science Centre (SSC) of the XMM-Newton satellite has recently reported results from an optical campaign aiming at the identification of the brightest X-ray sources in the XGPS \citep{motch2010}. The $\sim$\,3\,deg$^{2}$ area surveyed is located at $l$\,=\,20\degr, $b$\,=\,0\degr\ \citep{hands2004}. Among the 29 hard (2-10\,keV; S/N\,$\geq$\,3) sources investigated optically, six are identified with massive stars possibly containing an accreting component or being powered by colliding winds, three are identified with CVs, two with low-mass X-ray binary candidates and six with stars. At \Fx\ $\ga$  10$^{-13}$\ergscm\ (2-10\,keV), a large fraction of the expected Galactic source population is positively identified. Active coronae account for $\sim$\,10\% of the expected number of Galactic sources in the hard band.  

\section{2MASS and GLIMPSE identifications}
\begin{figure}[t!]
\resizebox{\hsize}{!}{\includegraphics[clip=true,angle=-90,bbllx=1.0cm,bburx=21.5cm,bblly=1cm,bbury=21.5cm]{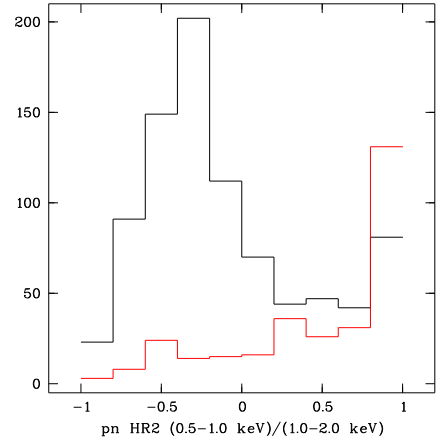}}
\caption{\footnotesize
Distribution of EPIC pn hardness ratio HR2 = ([1.0-2.0]\,--\,[0.5-1.0])/[0.5\,--\,2.0\,keV]) for $|b|<3$\degr\ 2XMM-DR3 sources with err(HR2)\,$\leq$\,0.1. Black: sources having a $\geq$\,90\% probability to be associated with a 2MASS entry; red: sources without any 2MASS entry within a combined X-ray + 2MASS 3$\sigma$ error radius.
}
\label{histo_pnhr2}
\end{figure}

\begin{figure}[t!]
\resizebox{\hsize}{!}{\includegraphics[clip=true,angle=-90,bbllx=1.0cm,bburx=21.5cm,bblly=1cm,bbury=21.5cm]{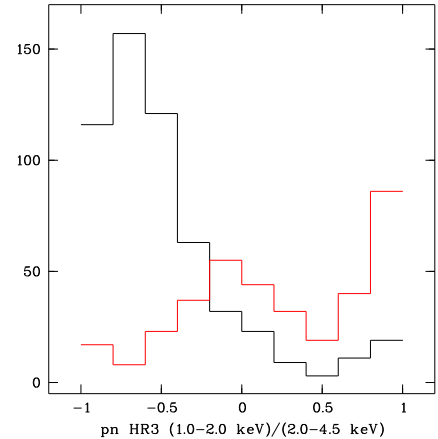}}
\caption{\footnotesize
Same as Fig.\ref{histo_pnhr2} for EPIC pn HR3 = ([2.0-4.5]\,--\,[1.0-2.0])/[1.0\,-\,4.5\,keV].}
\label{histo_pnhr3}
\end{figure}

The advent of high quality photometric infra-red surveys covering large areas offers a unique opportunity to statistically identify and characterise hard Galactic X-ray sources.  Using the method outlined in \cite{pineau2011}, we computed probabilities of identification with the 2MASS and GLIMPSE catalogues for all 2XMM-DR3 sources located within 3\degr\ from the Galactic plane. About one quarter of the $\sim$\,38,000 2XMM-DR3\footnote{The third release of the 2XMM catalogue was published in April 2010. see http://xmmssc-www.star.le.ac.uk/Catalogue/2XMMi-DR3/} low $b$ entries have a 2MASS match with an individual probability higher than 90\%. At this level, the expected number of spurious matches is $\sim$\,1.3\% \citep{motch2010}. Figs\,\ref{histo_pnhr2}\,and\,\ref{histo_pnhr3} show the distribution in EPIC pn hardness ratios of the 2XMM DR3 sources with and without 2MASS counterparts. Sources matching 2MASS entries are clearly much softer in X-rays. The peak in the HR2 histogram is consistent with optically thin thermal emission with kT\,$\sim$\,0.5\,keV undergoing an absorption with log\nh\,$\sim$\,21.5 while the HR3 distribution indicates the presence of a harder X-ray component (kT\,$\sim$\,1\,keV). Most 2MASS identifications are thus likely active coronae. The few soft XMM-Newton sources without 2MASS entries are likely Me stars being too faint in the infrared to be listed in the 2MASS catalogue. 

\begin{figure}[t!]
\resizebox{\hsize}{!}{\includegraphics[clip=true,angle=-90,bbllx=1.0cm,bburx=21.5cm,bblly=1cm,bbury=21.5cm]{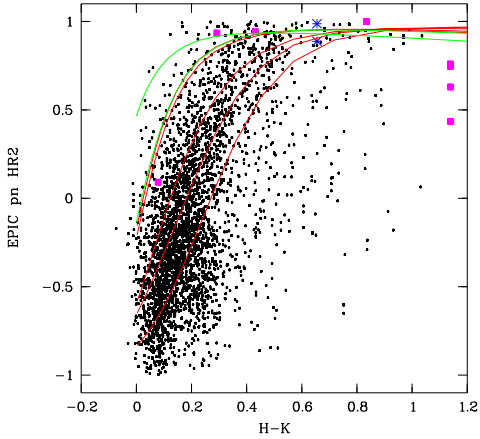}}
\caption{\footnotesize
H-K colour versus EPIC pn HR2 for all $|b|<3$\degr\ 2XMM-DR3 sources with err(HR2)\,$\leq$\,0.25, err(H-K)\,$\leq$\,0.1 and probability of 2MASS identification $\geq$\,90\%. The first three lower dark (red) lines show the expected E(H-K) versus HR2 relation assuming 2-T thermal emission from a 1.9\,Gyr, 300\,Myr and 30\,Myr active stars \citep{guedel1997}. The two hardest dark (red) relations correspond to AY Cet, a typical BY Dra binary \citep{dempsey1997} and to the RS CVn star WW Dra \citep{dempsey1993}. The two light (green) lines correspond to power laws with photon indices $\Gamma$ of 0 and 2. The RS CVn relation is almost superposed on the $\Gamma$\,=\,2 line. Positions of the WR star XGPS-14 \citep{motch2010}, (blue stars) and of a number of INTEGRAL HMXBs (big magenta squares) are also shown for comparison. Multiple detections are plotted.}       
\label{Th_k_pnhr2}
\end{figure}

\begin{figure}[t!]
\resizebox{\hsize}{!}{\includegraphics[clip=true,angle=-90,bbllx=1.0cm,bburx=21.5cm,bblly=1cm,bbury=21.5cm]{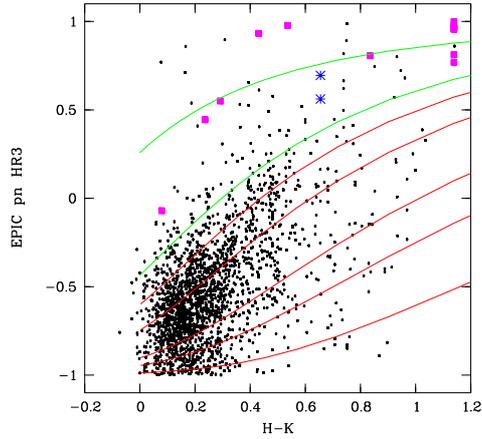}}
\caption{\footnotesize Same as Fig. \ref{Th_k_pnhr2} for EPIC pn HR3. WRs + INTEGRAL HMXBs are even better separated from coronal sources in this diagram.}
\label{Th_k_pnhr3}
\end{figure}

We show in Figs. \ref{Th_k_pnhr2} and \ref{Th_k_pnhr3} the distribution of EPIC pn HRs with the H-K colour. Since the intrinsic stellar H-K colour index remains within $-$0.1 to +0.1 from O to K spectral types and all luminosity classes \citep[see e.g.][]{covey2007}, the H-K colour mainly reflects interstellar absorption. No single X-ray energy distribution can account for the overall HR/E(H-K) relation. The most absorbed sources appear to be also the intrinsically hardest ones. The hotter X-ray temperature of young stellar coronae and active binaries combined with their higher luminosity make them detectable up to larger distances than older and X-ray softer stars. In particular, the bulk of the stars well detected above 2\,keV (i.e. appearing in Fig. \ref{Th_k_pnhr3}) have HR3 consistent with active binaries of the BY Dra or RS CVn type.


However, many very hard X-ray sources have high probability 2MASS identifications. Their number is significantly larger than expected from spurious matches. Assuming that their red H-K is due to interstellar absorption (\nh\,$\sim$\,10$^{22}$ cm$^{-2}$ for E(H-K)\,=\,0.3) yield distances larger than 3\,kpc (for a mean particle density of $n$\,=\,1). Sources located above the $\Gamma$\,=\,2 powerlaw curve in the H-K versus HR3 diagram shown in Fig. \ref{Th_k_pnhr3} have a H magnitude of about 11 and a H-K around 0.45. At a distance of 3\,kpc, the absolute H magnitude is $\leq$\,-1.7 suggesting that these 2MASS identifications could well be massive stars. Their positions in the HR/H-K diagram are indeed comparable to those of HMXBs discovered by INTEGRAL and are akin to the Wolf-Rayet (WR) XGPS-14 \citep{motch2010}. Therefore, based on the optical and infrared spectroscopic identification work reported by \cite{motch2010} and by \cite{anderson2011}, we expect that many of these hard X-ray 2MASS identifications are low-\Lx\ HMXBs, massive stars (either single or in wind colliding binaries) and $\gamma$-Cas analogs. 
Because of the high density of Spitzer sources in the Galactic plane, the probability to find at random a relatively bright GLIMPSE source in the XMM error circle is rather large. Only 222 2XMM-DR3 sources, for which boresight correction could be carried out \citep[see][]{watson2009}, have a matching probability higher than 90\%. To this we add 103 sources with P$\geq$ 90\% resulting from the crosscorrelation with XMM observations for which no boresight correction was possible. These sources have on average larger error circles \citep{watson2009}. XMM-GLIMPSE sources with a high probability of identification have a distribution in hardness ratios similar to that seen for 2MASS. Many of the bright XMM-GLIMPSE sources do have identifications with known optically bright active coronae. Among the hardest sources we find several catalogued WR stars, HMXBs discovered by INTEGRAL and a number of unidentified sources sharing very similar X-ray and infrared properties.    
\section{Hard X-ray sources with faint optical counterparts: - A Search for low \Lx\ accreting black holes}

Optically faint hard X-ray sources are much less constrained than those associated with luminous objects such as the massive stars discussed in the two previous sections. Deep infrared observations have shown indeed that the majority of the hard X-ray sources detected in the Galactic Centre region are not associated with massive stars \citep{laycock2005}. Optical follow-up of ChaMPlane sources in the Galactic bulge \citep{koenig2008} suggests the presence of a large population of active binaries and young stellar objects with a small CV contribution. Our ESO-VLT optical observations of the brightest hard X-ray sources in a region located at $l\,=\,20$\degr\, $b\,=\,0$\degr\, \citep{motch2010} led to the identification of five CVs or LMXB candidates and of a few active coronae. This distribution is globally consistent with the locally determined X-ray luminosity function of faint point sources reported by \cite{sazonov2006}. However, we expect the contribution of the various kinds of hard X-ray emitters to strongly vary with Galactic position. 

In addition, apart from toward the very central regions of the Galaxy, a large fraction of the hard X-ray sources detected at low Galactic latitude are background AGN. This extragalactic "contamination" depends sensitively on the direction of observation \citep{motch2006,hong2009}. A further difficulty arises from the fact that CV X-ray spectra resemble those of mildly absorbed AGN and in general cannot be efficiently preselected on the basis of hardness ratios only.

In order to increase our chances to select genuine Galactic X-ray sources for optical follow-up, we used the signature of the large photoelectric absorption imprinted on background AGN to achieve a high rejection rate for extragalactic sources. Our goals were twofold. First, investigate the nature of this optically faint hard X-ray Galactic population and second, constrain the surface density of low X-ray luminosity black holes (BH) in quiescent binaries, or being isolated and accreting from the interstellar medium. 

Apart from three cases of long micro-lensing events possibly due to isolated black holes \citep{bennett2002,nucita2006} and perhaps a couple of massive unseen companions in X-ray quiet binaries, all established or candidate stellar mass black holes ($\sim$\,60) are in accreting binaries \citep[see][for a recent census]{ziolkowski2010}. However, the actual number of isolated black holes (IBHs) in the Galaxy could be of the order of 10$^{8}$ \citep{samland1998,sartore2010} or even higher. A fraction of them might accrete matter from the interstellar medium at a rate high enough to become detectable in the radio, optical or X-ray domains. IBH X-ray properties have been investigated in details by \cite{agol2002}. Their detectability depends sensitively on three parameters. The first one is the assumed spatial velocity distribution, which reflects the amplitude of the kicks received at birth, and is critical for Bondi-Hoyle accretion. \cite{jonker2004} have shown that BH X-ray binaries display scale heights comparable to those of neutron star binaries.  This suggests that at birth BHs receive velocity kicks similar to those of neutron stars. The second key parameter is the efficiency of Bondi-Hoyle accretion, which can be quite significantly diminished in presence of magnetic fields. The accretion rate can be expressed as:
\begin{equation}
\dot{\rm M} = \lambda \ \frac{ 4 \pi G^{2} M^{2} \rho}{(v_{rel}^{2} + c_{s}^{2})^\frac{3}{2}} 
\end{equation}
with $v_{rel}$ the relative velocity of the accreting object with respect to the ISM having sound speed $c_{s}$ and mass density $\rho$. \cite{perna2003} argue that the value of the dimensionless parameter $\lambda$ which determines the actual efficiency of Bondi-Hoyle accretion, and is usually assumed to be of the order of unity, could be as low as 10$^{-2}$ or even less. The last important parameter is the efficiency $\epsilon$ with which X-rays are generated in the accretion flow (\Lx\,=\,$\epsilon \dot{\rm M} c^{2}$). The general agreement is that $\epsilon$ decreases sensitively at low $\dot{\rm M}$ due to the onset of radiatively inefficient accretion flows and as a result of an increasing fraction of accretion energy being transformed into kinetic energy of possible jets. 

\begin{figure}[t!]
\resizebox{\hsize}{!}{\includegraphics[clip=true,angle=-90,bbllx=1.0cm,bburx=21.5cm,bblly=1cm,bbury=21.5cm]{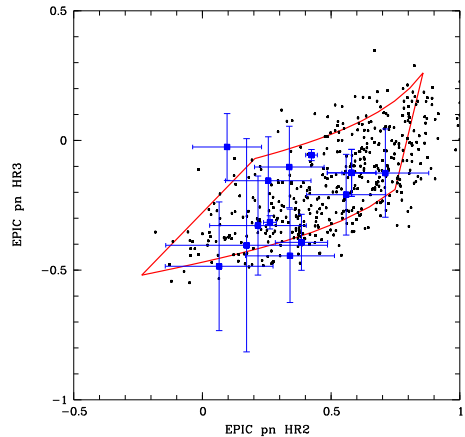}}
\caption{\footnotesize Distribution of candidate accreting black holes in the EPIC pn HR2/HR3 diagram (as defined in the 2XMM DR3 catalogue). The red lines define the area containing sources with X-ray energy distributions expected from accreting BHs undergoing \nh\ $\leq$ 10$^{22}$cm$^{-2}$. Candidates extracted from the 2XMM DR3 are shown as black dots. Sources selected for optical follow-up observations are plotted in blue together with their HRs errors.}
\label{plot_P75_etc}
\end{figure}

BH quiescent binaries display powerlaw X-ray spectra with photon index $\Gamma$ between 0.9 to 2.3 and X-ray luminosities in the range of 10$^{30}$ to 10$^{33}$\ergs\ \citep{kong2002,hameury2003}. Here we assume that IBHs exhibit similar X-ray spectra. As a best compromise between maximum distance range and survey area, we selected Galactic directions ($|b|$ $<$ 3\degr) toward which the total Galactic interstellar column is higher than 10$^{22}$cm$^{-2}$. For a typical ISM density of $n$\,$\sim$\,1\,cm$^{-3}$, this \nh\ is reached at distances of $\sim$\,3\,kpc. At the time we started this project we relied on observations contained in the 1XMM catalogue to which we added several pointings extracted from the XMM-Newton archive. In order to obtain consistent HR definitions, we only considered the EPIC pn camera. The total area fulfilling these conditions in our source selection was 20 \,deg$^{2}$ at \Fx\,$\geq$\,5\,10$^{-14}$\ergscm\ (0.2-12\,keV), taking into account overlapping exposures. This flux corresponds to a limiting X-ray luminosity of 6\,10$^{30}$\ergs\ at 1\,kpc. We then considered all sources displaying hardness ratios consistent within the errors with $\Gamma$\,=\,0.9--2.3 powerlaw energy distributions absorbed by \nh\,$\leq$\,10$^{22}$cm$^{-2}$. Among the latter, we selected for optical follow-up 14 of the X-ray brightest sources having optical candidates fainter than B\,=\,18\footnote{A fraction of the accretion luminosity should be emitted in the optical and infrared domain via synchrotron emission of the hot accreting plasma} and being observable from ESO La Silla during Chilean winter time. All selected X-ray sources were visually checked and we discarded areas of high diffuse X-ray and optical emission as well as star forming regions. The set of hardness ratios used in the 1XMM provides less energy resolution in the soft bands than those used subsequently. Nevertheless, the resulting 1XMM based source selection was found to be consistent with that based on 2XMM HRs as shown in Fig.\ref{plot_P75_etc}. Source selection and validation have been done using the XCat-DB\footnote{http://xcatdb.u-strasbg.fr/} \citep{motch2009}, the official SSC interface to XMM catalogues developed in Strasbourg.

Optical observations were carried out with the ESO-NTT + EMMI instrument from July 31 to August 2, 2005. All spectra were obtained with Grism \# 5 through a 1.5\arcsec\ slit. This setup provided a spectral resolution of 1000 in the wavelength range of 3800 to 7000\AA. 

We list in Tab. \ref{optids} a summary of the results of our optical identification work. Among these 14 sources we identified four cataclysmic variables, four Me stars and three active stars. 

\begin{table*}
\centering
\caption{Summary of optical identifications. AC\,=\,Active coronae earlier than M.}
\begin{minipage}{8.3cm}
\centering
\begin{tabular}{lccl}
\hline
\\
Source name           &  X-ray flux \footnote{In units of 10$^{-13}$\ergscm\ (0.2-12\,keV). Average of all detections.}  & V mag   & Identification \\
\\
\hline
\\
2XMM J135859.3-601518 & 1.87 & $>$ 22 &  UNID     \\  
2XMM J154305.5-522709 & 7.93 & 20.8   &  CV   \\
2XMM J172803.0-350039 & 10.6 & 18.5   &  Me     \\    
2XMM J174504.2-283552 & 0.58 & 17.5   &  AC  \\
2XMM J174512.9-290931 & 0.40 & 20.9   &  Me     \\    
2XMM J175520.5-261433 & 2.07 & $>$ 22 &  UNID   \\    
2XMM J180235.9-231332 & 2.87 & 20.0   &  CV  \\
2XMM J180243.0-224105 & 1.43 & 20.1   &  Me     \\    
2XMM J180913.0-190535 & 2.39 & 22.4   &  CV     \\
2XMM J181003.2-212336 & 8.24 & 16.2   &  AC    \\
2XMM J181857.9-160208 & 1.06 & $>$ 22 &  UNID   \\    
2XMM J182703.7-113713 & 0.65 & 17.3   &  AC   \\
2XMM J183228.1-102709 & 0.50 & 20.3   &  Me     \\    
2XMM J185233.2+000638 & 0.62 & 20.4   &  CV     \\
\\
\hline
\end{tabular}\par
   \vspace{-0.75\skip\footins}
   \renewcommand{\footnoterule}{}
  \end{minipage}
\label{optids}
\end{table*}

Only three XMM sources remain unidentified with candidate counterparts fainter than typically V\,=\,22.  We note however, that our optical spectroscopic limit is not yet deep enough to rule out an identification with a high \fxfo\ CV such as for instance 2XMM J183251.4-100106 (\Fx\,=\,8.5 10$^{-13}$\ergscm, V\,=\,23.2; \citealt{motch2010}). From this optical campaign, we derive an upper limit of 0.2 BH deg$^{-2}$ at \Fx\,=\,1.3\,10$^{-13}$\ergscm\ (0.2-12\,keV). This is $\ga$ 20 times the IBH surface density predicted by \cite{agol2002} for central regions of the Galaxy such as that covered by our optical targets ($b$\,=\,-1\degr,+2\degr\ and $l$\,=\,-49\degr,+33\degr). According to their model we expect 0.17 IBH in the total area surveyed at \Fx\,=\,1.3\,10$^{-13}$\ergscm\ and less than 3 at a 10 times lower flux limit. It should also be stressed that our limiting \nh\ de facto excludes from our sample all BH located at distances greater than a few kpc. This negative result clearly illustrates the difficulty of searching for Galactic IBHs in the X-ray domain where the background of AGN and the foreground of hard coronal emitters and CVs completely dominate the population. Using the population parameters of \cite{agol2002} our constraints can be expressed as $\lambda\,\times\,\epsilon\, \la\ 2\,10^{-4}\,/\,N_9 $ with $N_9$ the number of Galactic BHs in units of 10$^{9}$. For a reasonable value of $N_9$\,=\,0.2, we derive an upper limit of $\approx$\,10$^{-3}$ on the global (Bondi-Hoyle times X-ray) efficiency.

\section{Conclusions}

In this paper, we report on several efforts made to characterise the nature of the serendipitous XMM-Newton sources discovered in the Galactic plane. Dedicated optical follow-up observations and cross-correlation with archival catalogues reveal a significant number of hard X-ray emitting massive stars with luminosities in the range of $\sim$\,10$^{32}$ to $\sim$\,10$^{34}$\ergs. This population mainly consists of Wolf-Rayet stars, wind colliding binaries, HMXBs in quiescence and $\gamma$-Cas analogs. Young stars and active corona binaries of the BY Dra and RS CVn types also contribute significantly to the population of sources detected at energies above 2\,keV. The optically faintest Galactic hard X-ray sources are mostly identified with cataclysmic variables.  

The typical XMM-Newton sensitivity allows us to constrain the nature of the hard X-ray sources of low- to intermediate- \Lx \ up to distances of a few kpc. It is therefore unclear whether these studies can be used to assess the nature of the unresolved population responsible for the Galactic Ridge X-ray emission mostly seen at $|l|$ $\la$\,50\degr. 

Finally, we report on a first dedicated search for low-\Lx\ black holes in quiescent binaries or in isolation and accreting from the interstellar medium. We derive an upper limit of 0.2 BH deg$^{-2}$ at \Fx\,=\,1.3\,10$^{-13}$\ergscm\ (0.2-12\,keV) in directions of the central parts of the Galaxy. This observational limit implies that the efficiency $\epsilon$ with which X-rays are generated in the accretion flow is $\la$\,10$^{-3}$ if one assumes nominal Bondi-Hoyle mass accretion rates. 

\bibliographystyle{aa}

\bigskip
\bigskip
\noindent {\bf DISCUSSION}

\bigskip
\noindent {\bf MARAT GILFANOV:} Is there a similar on-going effort for the soft band ? 

\bigskip
\noindent {\bf CHRISTIAN MOTCH:} The Survey Science Centre of the XMM-Newton satellite has undertaken a wide area optical follow-up programme to identify serendipitous EPIC sources in different directions of the Galaxy. Most soft X-ray sources are identified with active coronae. The analysis of their X-ray and optical properties is one of the main scope of this identification programme.

\bigskip
\noindent {\bf TAKESHI GO TSURU:} The X-ray spectrum of the Galactic ridge emission is characterised by iron emission lines. Do you think if the collection of point sources can explain the ridge spectrum including iron emission lines ? 

\bigskip
\noindent {\bf CHRISTIAN MOTCH:} According to \citep{rev2009}, more than 80\% of the diffuse flux at energies of 6-7\,keV is resolved into discrete sources at $l$ = 0.08\degr, $b$= -1.42\degr. Therefore, it seems that the combination of faint stellar and CV sources responsible for the GRXE has intrinsic iron line flux strong enough to account for that seen in the unresolved emission.

\end{document}